\documentstyle[prl,twocolumn,aps]{revtex}
\begin{document}
\draft

\title{Universal diffusion near the golden chaos border}

\author {S.Ruffo$^{(a)}$, D.L.Shepelyansky$^{(b)}$}
\address {
Laboratoire de Physique Quantique, Universit\'{e} Paul Sabatier,\\
118, route de Narbonne, 31062 Toulouse, France
}
%
%                    *****************
%
%\date{preprint: LPHQTH 95/8 - NEIP 95-007}

\twocolumn[
\date{\today}
%\date{June 25, 1995}
\maketitle
\widetext

\vspace*{-1.0truecm}

\begin{abstract}
\begin{center}
\parbox{14cm}{We study local diffusion rate $D$ in Chirikov standard map near
the critical golden curve. Numerical simulations confirm the predicted
exponent $\alpha=5$ for the power law decay of $D$ as approaching the
golden curve via principal resonances with period $q_n$
($D \sim 1/q^{\alpha}_n$).
The universal self-similar structure of diffusion between
principal resonances is
demonstrated and it is shown that resonances of other type
play also an important
role.}
\end{center}
\end{abstract}
\pacs{
\hspace{1.9cm}
}
]
\narrowtext
During last years intensive investigations have allowed to understand
the structure of critical invariant curves at chaos border in
Hamiltonian dynamical systems with divided phase space~\cite{Green,Mackay}.
Usually the analysis
is carried out for two dimensional (2d) area preserving maps, the paradigm
being Chirikov standard map~\cite{Boris1}. The critical invariant curve
is characterized by the rotation number and its continued fraction expansion.
It has been shown that the small scale structure near invariant curves
with the same tail in this expansion is universal for all smooth Hamiltonians
with two degrees of freedom and for 2d maps~\cite{Green,Mackay,Escande}.
Among all invariant curves better studied are those with golden rotation
number $r_g= (\sqrt{5}-1)/2$, whose expansion is a series of $1$'s.
This $r_g$ is the most
irrational number and therefore it is believed that invariant curves with
golden tails are locally the most robust ones. The structure of the critical
golden curve has been studied by means of renormalization group approach
and it has been shown that the phase space structure is self-similar and
universal on small scales.

Different scaling exponents have been found
in this critical regime and they were successfully used to determine the
diffusion rate through the destroyed invariant curve called
cantorus~\cite{Boris1,Percival}. However, no flux passes through the golden
curve at the critical value of perturbation parameter $K=K_g$.
In this case the trajectory has only local diffusion rate $D$, which depends
on its distance $\Delta r_n$ to the golden curve.
This diffusion characterizes the
motion of a particle in the vicinity of invariant curve $r_g$ at different
levels $n$ of convergents $r_n=p_n/q_n$ of the continued fraction expansion of
$r_g$ ($r_1=1/1,r_2=2/3,r_3=3/5 \dots$).
One can expect a power law dependence of $D$ on the resonant approximant
$q_n$, namely $D \sim {q_n}^{-\alpha}$. Chirikov gave a simple argument
for $\alpha=5$~\cite{Boris2}. According to him $D \sim (\Delta r_n)^2 / t_n$,
where $\Delta r_n = \vert r_g - r_n \vert \sim {q_n}^{-2}$ and $t_n$ is the
typical inverse frequency $\Omega_n$ of small oscillations around
principal resonance $q_n$. Then, $\Omega_n \sim q_n \Delta \omega_n$, where
$\Delta \omega_n$ is the width of resonance $q_n$~\cite{Boris1}.
In the critical case Chirikov overlap criterion~\cite{Boris1} implies
$\Delta \omega_n \sim \Delta r_n$, which gives~\cite{Boris2}
\begin{equation}
D \approx A D_0/q_n^{5} \sim (\Delta r_n)^{5/2} \sim (\delta y_n)^\nu
\label{diffusion}
\end{equation}
where $D_0=K^2/(8\pi^2)$ is the quasilinear diffusion rate and
$\delta y_n = \vert y_n - y_g \vert$ is the distance
of unstable periodic orbit $y_n$ with rotation number $r_n$ to
the golden curve $y_g$ along the symmetry line.
The exponent $\nu = 2.14699 \dots$
can be determined from the exponent $\sigma$ for
$\delta y_n \sim 1/q_n^{\sigma}$ found in~\cite{Mackay} ($\nu = 5/\sigma$).

The fast decay of diffusion rate near the chaos border $r_g$ means
that a diffusing particle will never reach the border itself. This
slow diffusion gives a long sticking of trajectory around stable
islands on different renormalization levels. As a result the statistics
of Poincar\'e recurrences $P$ (integrated probability to return into a
given region
after a time larger than $\tau$) decays with $\tau$ as $P(\tau) \sim 1/\tau^p$.
Such kind of decay has been first observed in~\cite{Boris3}, where
the average value of the exponent $p \approx 1.5$ has been found.
Further investigations have shown that the power law decay of $P$
is a generic property of Hamiltonian systems with divided phase
space~\cite{Karney,Boris4}. However, according to numerical results
the power itself is not universal, varying in the range $1 < p < 2$.
Moreover different maps with golden chaos boundary give different
$p$ values~\cite{Boris4}, which seems to be
in contradiction with universal self-similar
structure of phase space near the golden border. Indeed renormalization
arguments give $p=3$~\cite{Ott}, which is in sharp contradiction with
numerical results. One of the possible reasons of the above contradiction is
sticking of particles near the stable islands {\it between} principal
resonances $q_n$. However, the attempts to take into account these intermediate
resonances gave $p=2$~\cite{Ott}, which is too large compared to the
numerical results for the golden border $1 < p < 1.35$~\cite{Boris4}.
Therefore the problem of Poincar\'e recurrences remains unsolved and more
detailed investigations of the phase space structure near the golden
border should be performed.
One of the reasons why the properties of $P(\tau)$ are so important is
that the correlation function of dynamical variables $C(\tau)$ and the
probability $\mu (\tau)$ to stay in a given region for a time $t  > \tau$
are related to $P(\tau)$
%2
\begin{equation}
C(\tau) \sim \mu(\tau) \sim \tau P(\tau)/< \tau > \sim 1/\tau^{p-1}
\label{ergodic}
\end{equation}
where $< \tau >$ is the average return time. The above relations follow
from the ergodicity of motion on the chaotic component of the phase
space~\cite{Boris3,Karney,Boris4}.
The decay of correlations with power $p_c=p-1 < 1$ can lead to a divergence
of global diffusion rate and to strong fluctuations in divided phase space.

In this Letter we investigate the behavior of local diffusion rate $D$ near
the critical golden curve. Our first aim was to verify the theoretical
prediction~(\ref{diffusion}) and to analyze the structure of $D$
at different renormalization levels. As a model we have chosen Chirikov
standard map
\begin{equation}
{\bar y} = y - K/(2 \pi) \sin (2\pi x) ~~,~~
{\bar x}= x + {\bar y}~~~\hbox{mod 1}
\label{stmap}
\end{equation}
with perturbation parameter corresponding to critical golden curve
$K=K_g=.97163540631 \dots$.
To measure the local diffusion rate $D=(\Delta y)^2/\Delta t$
we apply the efficient method used for the
investigations of Arnold and modulational diffusion in Ref.~\cite{Boris5}
($t$ is number of iterations).
This method
allows to measure very small diffusion rates (down to computer noise level)
with relatively small number of iterations. By this method,
we compute $D$ at different
resonant Fibonacci approximants $r_n=p_n/q_n$ of the critical curve $r_g$.
We use from $N_p=10$ to $N_p=100$ trajectories near
unstable periodic orbits of
period $q_n$ (these points had been determined by MacKay~\cite{Mackay}).
Each trajectory is integrated for about $T=1000 \times q_n$ iterations.
The total interval $T$ is divided into $N_w=10$ windows ,
where the average $y$ displacement
was computed with the smoothing function $f=\sin^{2 \beta}(\pi t N_w / T)$.
Usually
we take $\beta=4 ; 6$. Such smoothing allows to suppress regular oscillations
by a factor $\propto (N_w /T)^{4 \beta +3}$~\cite{Boris5}.
To control the accuracy of our numerical computation of $D$, we determine
$D$ also near stable periodic orbits in the center of resonance $q_n$,
where our
method gives a value $D \sim 10^{-34}$, which corresponds to the level of
computer round-off error in double precision.

Our results in Fig. 1 confirm the theoretical prediction~(\ref{diffusion})
for the variation of $D$ over more than 20 orders of magnitude.
The numerical fit gives $\alpha = 4.99 \pm 0.02$ and $A=0.0066$.
We attribute the fluctuations at small $q_n$ values to the fact that for
a large number of iterations trajectories can exit from the chaotic layer
corresponding to the initial $q_n$. This effect disappears for larger
$q_n$ where the local diffusion rate is sufficiently small
or for shorter number of iterations. Let us note that
the numerical value of $A$ is surprisingly
small. Our explanation for this fact is the following. According to
{}~\cite{Boris1} the action change after half period of rotation
in the chaotic separatrix layer is quite small
$\Delta y \sim 4 {\lambda}^2 \exp(-\pi \lambda/2)$
(here $\lambda=2\pi$ is the frequencies ratio).
This gives an order of magnitude estimate for
$A = D(q_n=1)/D_0 \sim (\Delta y)^2/(2D_0) \sim 0.003$ and
explains its small value. A more accurate estimate of $A$
requires taking into account higher orders of perturbation in $K$.
The measured diffusion rate in the chaotic component is well
separated from the diffusion in the stable regions
produced by numerical round-off errors (Fig.1).

The above result shows the global structure of the diffusion rate $D$
while approaching the chaos border via resonant approximants $q_n$.
However, an interesting question concerns the behavior of $D $
in between $r_n=p_n/q_n$ and $r_{n+1}$. The comparison of
$D$ on these scales should reflect the self-similar
structure of phase-space on different renormalization levels.
To check this self-similarity we measure $D$ on two symmetry lines
$x=0; 0.5$. The symmetry line $x=0.5$ crosses the main part of
chaotic layers and contains mainly unstable points, while
the other line $x=0$ passes mainly through stable islands.
The known structure of periodic orbits on symmetry lines of map~(\ref{stmap})
\cite{Mackay} allows to find the basic renormalization intervals
on stable $x=0$ and unstable $x=0.5$ lines.

For the unstable line the first renormalization level interval
is $\Delta y_1 =
\vert y_1 - y_7 \vert$ where $y_n$ is
the $y$ value of the unstable periodic orbit with rotation number
$r_n$ on the line $x=0.5$. The next interval
$\Delta y_2 = \vert y_4-y_{10} \vert $ lies
on the other side of the golden curve.
The $m-$th interval is $\Delta y_m = \vert y_{3m-2} - y_{3m+4} \vert$.
This $m-$series selects the subsequence of $n-$values
which we will denote by $n_m = 3m-2$.
Intervals with odd values of $m$ lie above the invariant curve $r_g$ and those
with even $m$ lie below. The self-similarity of the phase space implies
that the dependence of diffusion $D$
on the position inside $m-$th interval should be
approximately the same as in $m+1-$th interval
after ${q^5_{n_m}}$ rescaling. At larger
$m-$values the self-similarity is expected to become better and
better. To check numerically this self-similarity we
compute the diffusion rate at 320
homogeneously distributed points in the intervals $\Delta y_m$
for $m=1, \dots, 6$. To compare different intervals we rescale
the diffusion rate defining $D_R = {q^5_{n_{m+2}}} D_m$
where $D_m$ is the diffusion at $m-$th level. The position
inside each interval is denoted by
$\Delta y_R = \vert (y - y_{3m-2}) \vert / \Delta y_m$ for $y$ between
$y_{3m-2}$ and $y_{3m+4}$ ($0 \leq \Delta y_R \leq 1$).

The comparison of renormalized diffusion rate $D_R$ on two levels
$m=4$ and $m=6$ on the unstable line is shown in Fig. 2. The diffusion rate
is self-similar in agreement with universal phase space structure near
the golden curve. The minimal diffusion rate $D_R$ on levels $m=4;6$ is
determined by computer noise and is different on the two levels
due to the different
normalization factors. The rare fluctuations in the upper diffusion
plateau are presumably due to exits of trajectory from initial chaotic layer.
Points with high diffusion rate are in the chaotic component, while
those with minimal diffusion correspond to trajectories in stable islands.
The self-similarity was also observed at other $m$ levels, both when
intervals were on the same side or on opposite sides of the golden curve.
This self-similarity becomes better with the growth of $m$.
The sharp separation between two levels of diffusion allows
to determine the leading resonances in each renormalization interval.
The biggest gap in the diffusion is for periodic orbit (labelled {\it b} in
Fig. 2) with rotation number $r_b=[{\{111\}}_{m-1},1,2,1]$, where
the triplet of $1$'s in curly brackets is repeated $m-1$ times. This rotation
number is not from the series of principal resonances given by the continued
fraction expansion of $r_g$ and labelled by {\it i} in Fig. 2
($r_i=[{\{111\}}_{m-1},1,1,1,1,1]$). While resonance $r_b$ is significantly
larger than the principal one $r_i$, there are also other resonances which are
of comparable or smaller size than $r_i$: $r_c=[{\{111\}}_{m-1},1,2,1,1,1,1]$,
$r_d=[{\{111\}}_{m-1},1,2,2,2,1]$, $r_e= [{\{111\}}_{m-1},1,2,2,1]$,
$r_g=[{\{111\}}_{m-1},1,1,1,3,2,1]$,
$r_h=[{\{111\}}_{m-1},1,1,1,\- 2,1,1,1]$,
$r_l=[{\{111\}}_{m-1},1,1,1,1,1,2,1,1]$. It is interesting to remark
that most of these rotation numbers have continued fraction expansions
containing mainly $1$'s and $2$'s, which is in qualitative agreement with
a conjecture made in ~\cite{Boris4} on the basis
of analysis of the Schmidt-Bialek
fractal diagram of map~(\ref{stmap}). This fact is also in agreement with
numerical results in~\cite{Stark}.

In Fig. 2 we see also other types of
resonances, namely $r_a=p_a/q_a=[{\{111\}}_{m-1},1,4,1]$ and $r_f=
p_f/q_f=[{\{111\}}_{m-1},1,1,1,3,1]$.
However, they have a different structure than previous ones, displaying
two and three intersections with unstable line $x=0.5$ giving
rise to double ({\it a}) and triple ({\it f}) drops in diffusion.
Indeed the
periodic orbit inside resonance $r_a$ ($r_f$) has period $2 q_a$ ($3 q_f$).
These orbits are not present in the limit $K \to 0$, and correspond to a new
chain of islands in the chaotic layer around resonances $r_a$ ($r_f$).
The important consequence of the analysis of the structure of these
resonances is that, in spite of the fact that renormalization group describes
quite well the convergence to golden curve, we see that on each renormalization
level resonances other than those of the main series of $r_g$, and even
some which are not present at $K=0$, occupy a sizeable part of phase space.
The existence of such non-standard resonances might explain the lack of
universality of the exponent $p$ in ~(\ref{ergodic}) and the numerical
value of $p$ significantly less than $2$. Indeed, very long Poincar\'e
recurrences can be originated by sticking of orbits not only near the
main resonances $r_n$ but also around these non-standard resonances and
the chains of islands around them. The general description of the phase-space
structure should take into account the presence of these resonances. An
interesting question is which is the weakest cantorus on a
given renormalization
level which will determine the transition time
between different renormalization
levels. According to Fig.2 all drops in $D$ are associated to periodic orbits
and not to invariant curves. This means that at our accuracy level there
is no other invariant curve except $r_g$.

We have also studied the local diffusion on the stable line $x=0.0$ (Fig. 3).
The size of renormalization interval $n$ is defined
as $\Delta y_n= \vert y_n - y_{n+2} \vert $ and the variable $\Delta y_R=
\vert (y - y_{n}) \vert / \Delta y_n$.
The structure of diffusion is also self-similar. As expected, the main
part of the renormalization interval is occupied by stable islands. The largest
resonances with rotation number $\rho_a=[{\{1\}}_{n}]$ and
$\rho_i=[{\{1\}}_{n},1,1]$ correspond to the main
series of $r_g$. There are also other resonances: $\rho_c=[{\{1\}}_{n},3,1]$,
$\rho_d=[{\{1\}}_{n},3,1,1]$, $\rho_e=[{\{1\}}_{n},2,1]$,
$\rho_f=[{\{1\}}_{n},2,1,1,1]$, $\rho_g=[{\{1\}}_{n},2,1,1]$,
$\rho_h=[{\{1\}}_{n},2,2,1]$. Again the biggest resonances have only $1$'s
and $2$'s in the continued fraction. Some of the resonances seen on $x=0.0$
are also observed on $x=0.5$ and one can easily establish the correspondence.
The resonance $\rho_b$ has the same rotation number as $\rho_a$,
but it corresponds to
a different orbit. The period of this orbit is $9$
times larger than that of $\rho_a$ and
the rotation number around main islands of $\rho_a$ is $1/9$, that is
not in the golden mean sequence. This orbit is located in the center of
stable islands which are embedded
in the chaotic layer around principal resonances.
This orbit also does not exist at $K=0$.
Generally, the analysis performed on the two symmetry lines confirms
the self-similar structure of diffusion
and demonstrates the important role of non-standard resonances.

The above analysis allows to understand some important properties of
local diffusion rate $D$ near the critical invariant curve $r_g$.
The self-similar structure of $D$ shows also the importance of non-standard
resonances, different from principal approximants $r_n=p_n/q_n$ of
$r_g$. These non-standard resonances are also self-similar at different
renormalization levels. However, their sizes in phase space are comparable
or sometimes even larger than those of principal resonances.
Therefore trajectories can be trapped for a long time around these
non-standard resonances and diffuse very slowly to internal chaos boundaries
surrounding islands of these resonances. Since the sizes of non-standard
and principal resonances are comparable, the contribution of internal
chaos boundaries to Poincar\'e recurrences may be relevant, and thus
the decay of $P(\tau)$ may be non universal.
The determination of the asymptotic behavior of
$P(\tau)$ requires also a better
understanding of transition rates between different renormalization levels,
which are not directly related to local diffusion and should be studied in more
detail.

One of us (S.R.) gratefully acknowledges the hospitality and financial
support of the Laboratoire de Physique Quantique, URA 505, CNRS at
Universit\'e Paul Sabatier, Toulouse, France. This work is also
part of European contract \# ERBCHRXCT940460.

\begin{figure}
\caption{Local diffusion rate $D$ near principal resonance
convergents $r_n=p_n/q_n$ to $r_g=(5^{1/2} -1)/2$, $D_0 = K^2/(8 \pi^2)$;
chaotic component near
unstable points with {\it a)} $T=10^{3} q_n$, $N_w=10$, $N_p=10$ ($\times$);
{\it b)}$T=10^{4} q_n$,
$N_w=100$, $N_p=100$ ($\bigcirc$);
{\it c)}regular component near stable points with
$T=10^{3} q_n$, $N_w=10$, $N_p=10$ (triangles). The straight line shows the
theoretical slope $\alpha=5$. }

\label{fig1}
\end{figure}

\begin{figure}
\caption{Renormalized diffusion rate $D_R/D_0$ {\it vs.} rescaled position
$\Delta y_R$ in the $m-$th level interval of renormalization scheme on the
unstable symmetry line $x=0.5$;
$m=4$ (full curve) and $m=6$ (dashed line with crosses). The letters
indicate the drops of diffusion $D_R$ (see text). Other parameters are as in
Fig. 1 {\it a)}.
}
\label{fig2}
\end{figure}

\begin{figure}
\caption{Renormalized diffusion rate $D_R/D_0$ {\it vs.} rescaled position
$\Delta y_R$ in the $n-$th level interval of renormalization scheme on the
stable symmetry line $x=0.0$;
$n=8$ (full curve) and $n=10$ (dashed line with crosses). The letters
indicate the drops of diffusion $D_R$ (see text); $T=2 \times10^{4} q_n$,
$N_w=10$, $N_p=10$.}
\label{fig3}
\end{figure}

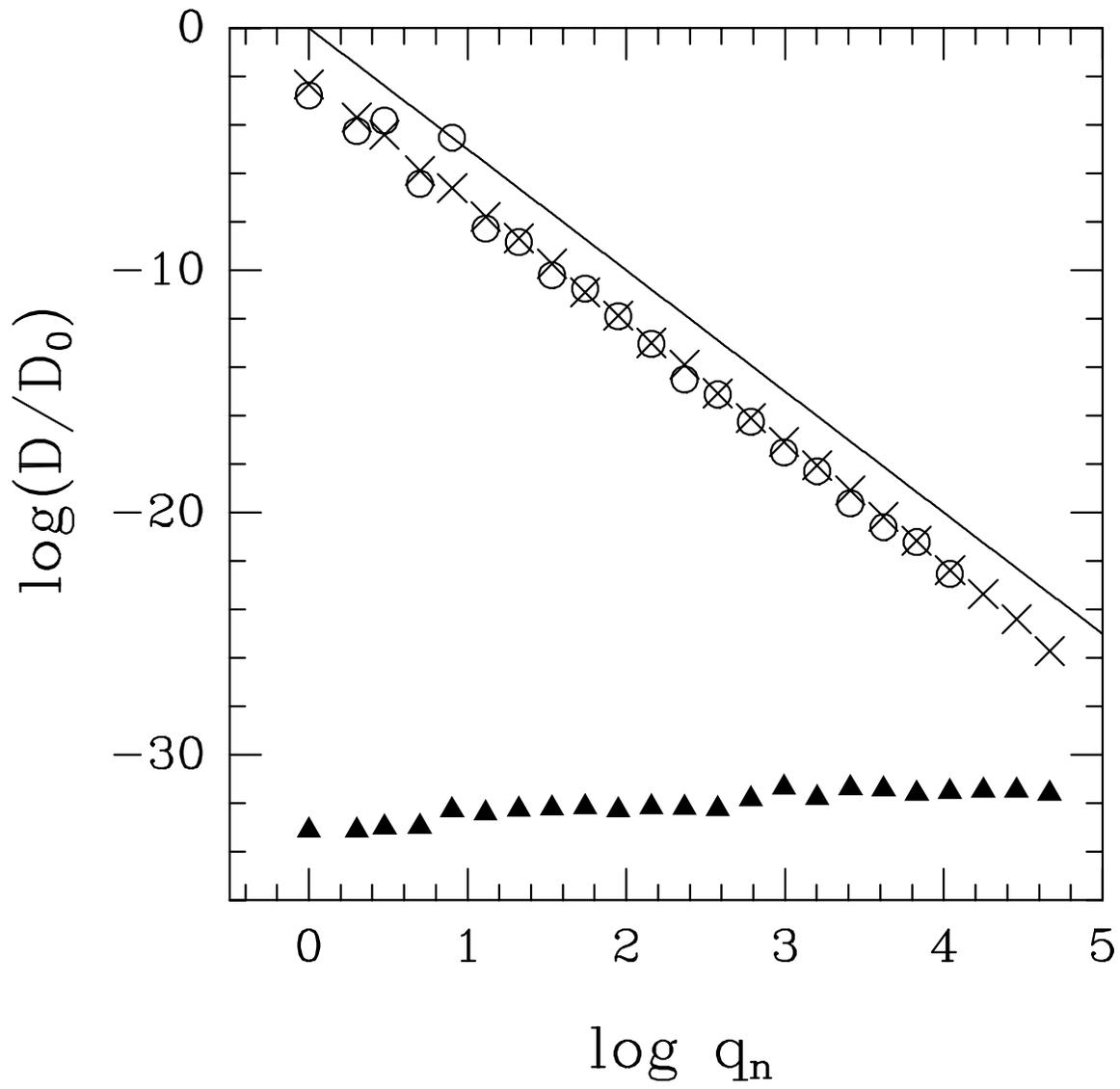
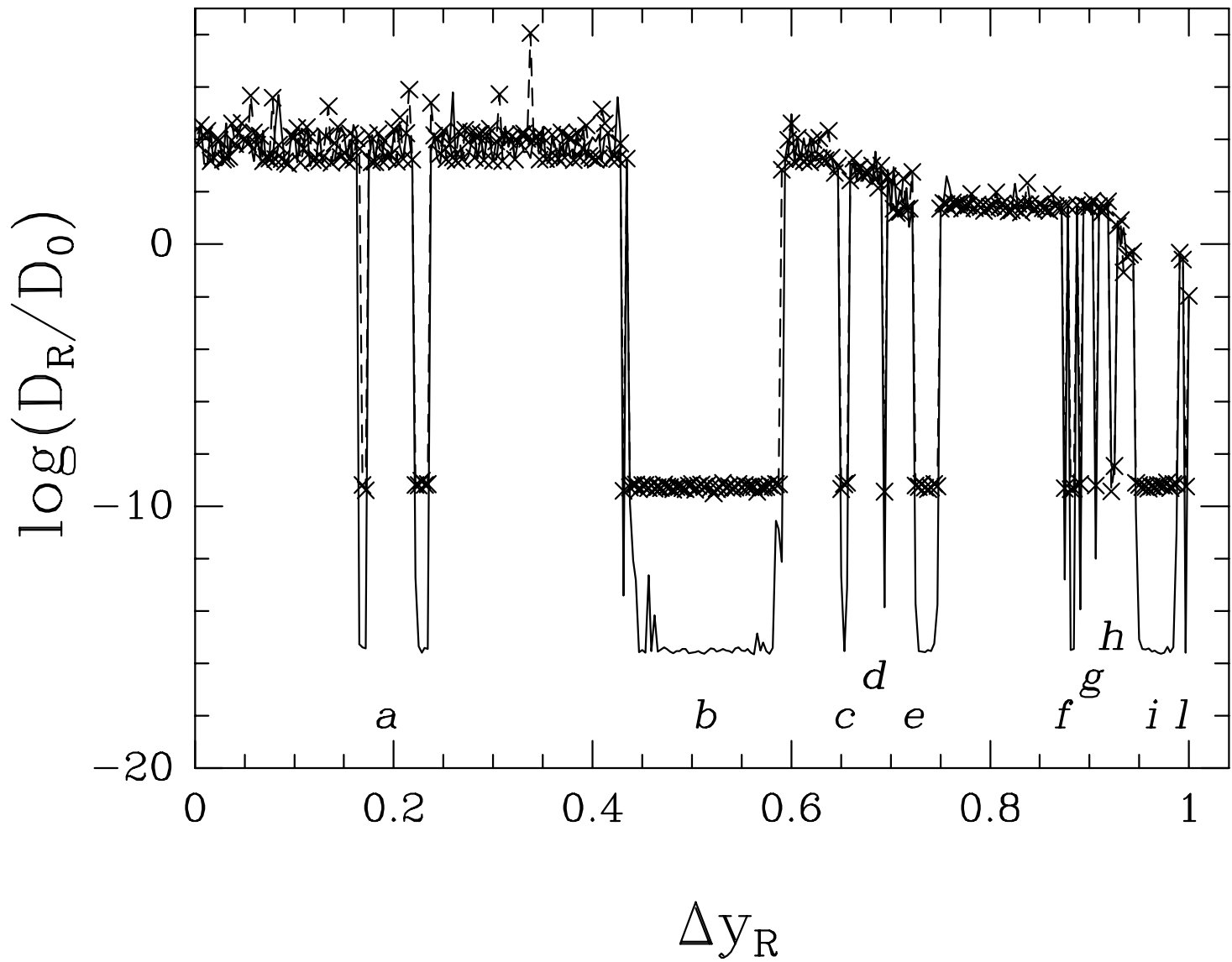
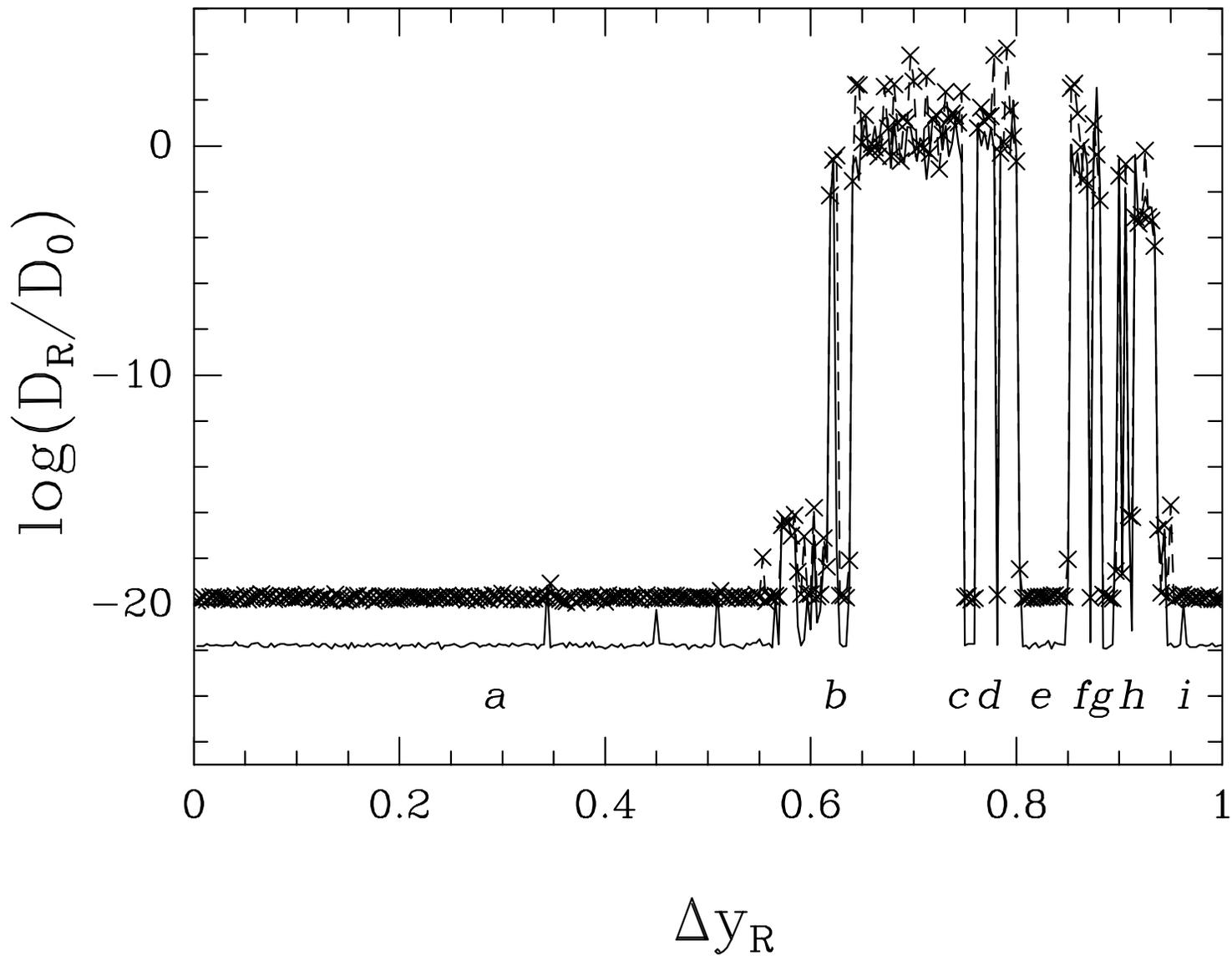
\end{document}